\documentclass[a4paper,UKenglish]{lipics-v2018}
\usepackage{microtype,mathdefs,array,amsthm,enum2,relsize,booktabs}
\usepackage[section]{thmdefs}

\let\sc\scshape

\let\ifunfinished\iftrue

\makeatletter
\newcommand\m@thsm@ller[2]{\mbox{\relscale{0.91}$\m@th#1#2$}}
\let\smaller\undefined
\DeclareRobustCommand\smaller[1]{\relax\ifmmode{\mathpalette\m@thsm@ller{#1}}\else{\relscale{0.91}#1}\fi}
\makeatother

\DeclareRobustCommand*{\dom}{\qopname\relax o{dom}}

\newcommand*{\bisim}{\mathrel{\,\sim\,}}

\newcommand*{\tp}{\mathrm{tp}}
\newcommand*{\TP}{\mathrm{TP}}

\newcommand*{\Th}{\mathrm{Th}}
\renewcommand*{\TH}{\mathrm{TH}}

\newcommand*{\MSO}{\smaller{\mathrm{MSO}}}
\newcommand*{\CTL}{\smaller{\mathrm{CTL}}}
\newcommand*{\Lmu}{\mathrm L_\mu}
\newcommand*{\fuse}{\mathrm{fuse}}

\newcommand*{\stp}{\mathrm{stp}}
\newcommand*{\wtp}{\mathrm{wtp}}

\newcommand*{\?}{\kern .08em}

\newcommand\upqed{\vskip-\baselineskip\vskip-\belowdisplayskip}
\newcommand\markenddef{\hfill$\lrcorner$}

\title{Bisimulation Invariant\newline Monadic-Second Order Logic in the Finite}
\author{Achim Blumensath}{Masaryk University Brno}{blumens@fi.muni.cz}{}{Work supported by the Czech Science Foundation, grant No.~GA17-01035S}
\author{Felix Wolf}{Technische Universit\"at Darmstadt, Institute TEMF\\
Graduate School of Excellence Computational Engineering}{wolf@temf.tu-darmstadt.de}{}{Work partially supported by the \emph{Excellence Initiative} of the
German Federal and State Governments and the Graduate School of Computational
Engineering at Technische Universit\"at Darmstadt}
\authorrunning{A. Blumensath, F. Wolf}
\titlerunning{Bisimulation Invariant MSO in the Finite}
\Copyright{Achim Blumensath, Felix Wolf}
\ArticleNo{xx}

\begin{document}
\nolinenumbers
\maketitle

\begin{abstract}
We consider bisim\-u\-la\-tion-invari\-ant monadic second-order logic over various
classes of finite transition systems. We present several combinatorial characterisations
of when the expressive power of this fragment coincides with that of the modal
$\mu$-calculus. Using these characterisations we prove for some simple classes of
transition systems that this is indeed the case.
In particular, we show that,
over the class of all finite transition systems with Cantor--Bendixson rank at most~$k$,
bisim\-u\-la\-tion-invari\-ant $\MSO$ coincides with~$\Lmu$.
\end{abstract}

\section{Introduction}   %%%%%%%%%%%%%%%%%%%%%%%%%%%%%%%%%%%%%%%%%%%%%%%%%%%%%%%%%%%%%%%%%

A characterisation of the bisim\-u\-la\-tion-invari\-ant fragment of a given
classical logic relates this logic to a suitable modal logic.
In this way, one obtains a correspondence between a family of classical logics
and a family of modal logics.
Such characterisation results therefore help with ordering the zoo of logics
introduced (on both sides) over the years and with distinguishing between
natural and artificial instances of such logics.

The study of bisim\-u\-la\-tion-invari\-ant fragments of classical logics
was initiated by a result of van Benthem~\cite{vanBenthem76}
who proved that the bisim\-u\-la\-tion-invari\-ant fragment of first-order logic
coincides with standard modal logic.
Inspired by this work, several other characterisations have been obtained,
the most prominent among them being a characterisation of bisim\-u\-la\-tion-invari\-ant
monadic second-order logic by Janin and Walukiewicz~\cite{JaninWalukiewicz96}.
The table below summarises the results known so far.

\medskip
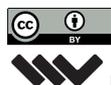
{\centering
\begin{tabular}{lll}
\toprule
  bisimulation-invariant fragment & modal logic & reference \\
\midrule
  first-order logic               & modal logic               & \cite{vanBenthem76} \\
  monadic second-order logic      & modal $\mu$-calculus      &
    \cite{JaninWalukiewicz96} \\
  monadic path logic              & $\CTL^*$                  &
    \cite{MollerRabinovitch99,MollerRabinovitch03} \\
  weak monadic second-order logic & continuous $\mu$-calculus & \cite{Carreiro15} \\
  weak chain logic                & $\mathrm{PDL}$            & \cite{Carreiro15} \\
\bottomrule
\end{tabular}\par}

\bigskip
There are also similar characterisations for various variants of bisimulation
like \emph{guarded bisimulation} \cite{AndrekaVBeNe98,GraedelHirschOtto02}
or bisimulation for \emph{inquisitive modal logic}~\cite{CiardelliOtto17}.

\medskip
Researchers in finite model theory started to investigate to which extent these
correspondences also hold when only considering \emph{finite} structures, that is,
whether every formula of a given classical logic that is bisim\-u\-la\-tion-invari\-ant
over the class of all finite transition systems is equivalent, over that class,
to the corresponding modal logic.
For first-order logic, a corresponding characterisation does indeed hold.
Its proof by Rosen~\cite{Rosen97} uses tools from finite model theory and
is very different to the proof by van Benthem.

The above mentioned result by Janin and Walukiewicz on bisim\-u\-la\-tion-invari\-ant
monadic second-order logic has so far defied all attempts at a similar transfer to
the realm of finite structures. The main reason is that the original proof is
based on automata-theoretic techniques and an essential ingredient is a reduction to
trees, via the unravelling operation. As this operation produces infinite trees,
we cannot use it for formulae that are only bisim\-u\-la\-tion-invari\-ant
over finite transition systems.

In this paper we start a fresh attempt at a finitary version of the result of Janin
and Walukiewicz. Instead of automata-theoretic techniques we employ the composition
method. For certain classes of very simple, finite transition systems we characterise
the bisim\-u\-la\-tion-invari\-ant fragments of monadic second-order logic over these
classes.
We hope that some day our techniques can be extended to the general case of \emph{all}
finite structures, but currently there are still a few technical obstacles to overcome.

We start in Section~\ref{Sect: bisim-inv} by recalling the needed material on
bisimulation and by listing all known results on bisim\-u\-la\-tion-invari\-ant
monadic second-order logic.
We also collect some low-hanging fruit by proving two new results concerning
(i)~finite classes and (ii)~the class of all finite trees.
Finally, we lay the groundwork for the more involved proofs to follow
by characterising bisim\-u\-la\-tion-invari\-ance in terms of a combinatorial
property called the \emph{unravelling property.}
In Section~\ref{Sect: composition}, we collect some tools from logic we will need.
The emphasis in on so-called \emph{composition lemmas.} Nothing in this section is new.

Finally we start in Section~\ref{Sect: types} in earnest by developing the technical
machinery our proofs are based on.
Sections \ref{Sect: lassos}~and~\ref{Sect: hierarchical lassos} contain our first
two applications\?: characterisations of bisim\-u\-la\-tion-invari\-ant monadic
second-order logic over (i)~the class of lassos and (ii)~certain classes of
what we call \emph{hierarchical} lassos.
The former is already known and simply serves as an example of our techniques
and to fix our notation for the second result, which is new.

Before presenting our last characterisation result,
we develop in Section~\ref{Sect: reductions} some additional technical tools
that allow us to reduce one characterisation result to another.
This is then applied in Section~\ref{Sect: CB rank} to the most complex of our results.
We characterise bisim\-u\-la\-tion-invari\-ant monadic second-order
logic over the class of all transition systems of a given Cantor--Bendixson rank.

\section{Bisimulation-invariance}   %%%%%%%%%%%%%%%%%%%%%%%%%%%%%%%%%%%%%%%%%%%%%%%%%%%%%%%%%%%%%%%%
\label{Sect: bisim-inv}

We consider two logics in this paper\?:
(i) \emph{monadic second-order logic} ($\MSO$), which is the extension of first-order logic
by set variables and set quantifiers, and (ii) the \emph{modal $\mu$-calculus} ($\Lmu$),
which is the fixed-point extension of modal logic.
A~detailed introduction can be found, e.g., in~\cite{GraedelThomasWilke02}.
Concerning the $\mu$-calculus and bisimulation, we also refer to the
survey~\cite{Stirling11}.
\emph{Transition systems} are directed graphs where the edges are labelled by
elements of a given set~$A$ and vertices by elements of some set~$I$.
Formally, we consider a transition system as a structure of the form
$\frakS = \langle S, (E_a)_{a \in A},(P_i)_{i \in I},s_0\rangle$
where the $E_a \subseteq S \times S$ are (disjoint) binary edge relations,
the $P_i \subseteq S$ are (disjoint) unary predicates, and $s_0$~is the initial state.
We write $\frakS,s$ to denote the transition system obtained from~$\frakS$
by declaring~$s$ to be the initial state.

A central notion in modal logic is \emph{bisimilarity}
since modal logics cannot distinguish between bisimilar systems.
\begin{Def}
Let $\frakS$~and~$\frakT$ be transition systems.

(a) A \emph{bisimulation} between $\frakS$~and~$\frakT$
is a binary relation $Z \subseteq S \times T$ such that all pairs
$\langle s,t\rangle \in Z$ satisfy the following conditions.
\begin{description}
\item[\normalfont (prop)] $s \in P^\frakS_i \ \iff\ t \in P^\frakT_i$,\quad
  for all $i \in I$.
\item[\normalfont (forth)] For each edge $\langle s,s'\rangle \in E^\frakS_a$,
  there is some $\langle t,t'\rangle \in E^\frakT_a$ such that
  $\langle s',t'\rangle \in Z$.
\item[\normalfont (back)] For each edge $\langle t,t'\rangle \in E^\frakT_a$,
  there is some $\langle s,s'\rangle \in E^\frakS_a$ such that
  $\langle s',t'\rangle \in Z$.
\end{description}

(b) Let $s_0$~and~$t_0$ be the initial states of, respectively,
$\frakS$~and~$\frakT$. We say that $\frakS$ and $\frakT$ are
\emph{bisimilar} if there exists a bisimulation~$Z$ between
$\frakS$~and~$\frakT$ with $\langle s_0,t_0\rangle \in Z$.
We denote this fact by $\frakS \bisim \frakT$.

(c) We denote by $\calU(\frakS)$ the \emph{unravelling} of a
transition system~$\frakS$.
\markenddef
\end{Def}

The next two observations show that the unravelling operation is closely related
to bisimilarity.
In fact, having the same unravelling can be seen as a poor man's version
of bisimilarity.
\begin{Lem}\label{Lem: unravelling preserves bisimulation}
Let\/ $\frakS$~and\/~$\frakT$ be transition systems.
\begin{enuma}
\item $\calU(\frakS) \bisim \frakS\,.$
\item $\frakS \bisim \frakT \qtextq{implies} \calU(\frakS) \bisim \calU(\frakT)\,.$
\end{enuma}
\end{Lem}
\begin{proof}
For (a), note that graph of the canonical homomorphism $\calU(\frakS) \to \frakS$
forms a bisimulation.
(b) follows by (a) since
$\calU(\frakS) \bisim \frakS \bisim \frakT \bisim \calU(\frakT)\,.$
\end{proof}

As already mentioned modal logics cannot distinguish between bisimilar systems.
They are \emph{bisim\-u\-la\-tion-invari\-ant} in the sense of the following definition.
\begin{Def}
Let $\calC$~be a class of transition systems.

(a) An $\MSO$-formula~$\varphi$ is \emph{bisim\-u\-la\-tion-invari\-ant}
over~$\calC$ if
\begin{align*}
  \frakS \sim \frakT
  \qtextq{implies}
  \frakS \models \varphi \ \Leftrightarrow\ \frakT \models \varphi\,,
  \quad\text{for all } \frakS,\frakT \in \calC\,.
\end{align*}

(b)
We say that, \emph{over the class~$\calC$, bisim\-u\-la\-tion-invari\-ant $\MSO$
coincides with~$\Lmu$} if, for every $\MSO$-formula~$\varphi$ that is
bisim\-u\-la\-tion-invari\-ant over the class~$\calC$, there exists an
$\Lmu$-formula~$\psi$ such that
\begin{align*}
  \frakS \models \varphi
  \quad\iff\quad
  \frakS \models \psi\,,
  \quad\text{for all } \frakS \in \calC\,.
\end{align*}
\upqed\markenddef
\end{Def}

A straightforward induction over the structure of formulae shows that every
$\Lmu$-formula is bisim\-u\-la\-tion-invari\-ant over all transition systems.
Hence, bisim\-u\-la\-tion-invari\-ance is a necessary condition for an $\MSO$-formula
to be equivalent to an $\Lmu$-formula.

The following characterisations of bisim\-u\-la\-tion-invari\-ant $\MSO$ have been
obtained so far. We start with the result of Janin and Walukiewicz.
\begin{Thm}[Janin, Walukiewicz \cite{JaninWalukiewicz96}]
Over the class of all transition systems,
bisim\-u\-la\-tion-invari\-ant $\MSO$ coincides with~$\Lmu$.
\end{Thm}

The main step in this theorem's proof consists in proving the following variant,
which implies the case of all structures by a simple reduction.
\begin{Thm}[Janin, Walukiewicz]\label{Thm: bisim-inv MSO over trees}\label{Thm: bisim MSO over trees}
Over the class of all trees, bisim\-u\-la\-tion-invari\-ant $\MSO$ coincides with~$\Lmu$.
\end{Thm}

There have already been two attempts at a finitary version.
The first one is by Hirsch who considered the class of all regular trees, i.e.,
unravellings of finite transition systems.
The proof is based on the fact that a formula is bisim\-u\-la\-tion-invari\-ant over
all trees if, and only if, it is bisim\-u\-la\-tion-invari\-ant over regular trees.
\begin{Thm}[Hirsch \cite{Hirsch02}]
Over the class of all regular trees, bisim\-u\-la\-tion-invari\-ant $\MSO$ coincides
with~$\Lmu$.
\end{Thm}

The second result is by Dawar and Janin who considered the class of finite lassos,
i.e., finite paths leading to a cycle.
We will present a proof in Section~\ref{Sect: lassos} below.
\begin{Thm}[Dawar, Janin \cite{DawarJanin04}]
Over the class of all lassos, bisim\-u\-la\-tion-invari\-ant $\MSO$ coincides with~$\Lmu$.
\end{Thm}

\smallskip
In this paper, we will extend this last result to larger classes.
We start with two easy observations.
The first one is nearly trivial.
\begin{Thm}\label{Thm: finite classes}
Over every finite class~$\calC$ of finite transition systems,
bisim\-u\-la\-tion-invari\-ant $\MSO$ coincides with~$\Lmu$.
\end{Thm}
\begin{proof}
As any two non-bisimilar, finite transition systems can be distinguished by an
$\Lmu$-formula (in fact, even by a formula of modal logic, see e.g.~\cite{Stirling11}),
we can pick, for every pair of non-bisimilar transition systems $\frakS,\frakT \in \calC$,
an $\Lmu$-formula satisfied by~$\frakS$, but not by~$\frakT$.
Let $\Theta$~be the resulting set of formulae.
The \emph{$\Theta$-theory} of a transition system $\frakS \in \calC$ is
\begin{align*}
  T_\Theta(\frakS) := \set{ \vartheta \in \Theta }{ \frakS \models \vartheta }\,.
\end{align*}
By choice of~$\Theta$ it follows that
\begin{align*}
  \frakT \models \Land T_\Theta(\frakS)
  \quad\iff\quad \frakT \bisim \frakS\,,
  \quad\text{for } \frakS,\frakT \in \calC\,.
\end{align*}

Given an $\MSO$-formula~$\varphi$ that is bisim\-u\-la\-tion-invari\-ant over~$\calC$,
we set
\begin{align*}
  \psi := \Lor {\bigset{ {\textstyle\Land T_\Theta(\frakS)} }
                       { \frakS \in \calC\,,\ \frakS \models \varphi }}\,.
\end{align*}
(As $\Theta$~is finite, this is a finite disjunction of finite conjunctions.)
Then $\psi \in \Lmu$ and, for each $\frakS \in \calC$, it follows that
\begin{align*}
  \frakS \models \psi
  \quad\iff\quad \frakS \bisim \frakT
    \quad\text{for some } \frakT \in \calC \text{ with } \frakT \models \varphi
  \quad\iff\quad \frakS \models \varphi\,.
\end{align*}
\upqed
\end{proof}

The second observation is much deeper, but fortunately nearly all of the work has already
been done by Janin and Walukiewicz.
\begin{Thm}\label{Thm: finite trees}
Over the class of all finite trees,
bisim\-u\-la\-tion-invari\-ant $\MSO$ coincides with~$\Lmu$.
\end{Thm}
\begin{proof}
We adapt the proof of Janin and Walukiewicz~\cite{JaninWalukiewicz96}
which roughly goes as follows.
For a transition system~$\frakM$, let $\widehat\frakM$~be the tree obtained
from the unravelling $\calU(\frakM)$ by duplicating every subtree infinitely many times.
Given an $\MSO$-formula~$\varphi$, one can use automaton-theoretic techniques
to construct an $\Lmu$-formula~$\varphi^\lor$ such that
\begin{align*}
  \widehat\frakM \models \varphi \quad\iff\quad \frakM \models \varphi^\lor\,.
\end{align*}
This is the contents of Lemma~12 of~\cite{JaninWalukiewicz96}.
Now the claim follows by bisim\-u\-la\-tion-invari\-ance since
\begin{align*}
  \frakM \models \varphi^\lor
  \quad\iff\quad \widehat\frakM \models \varphi
  \quad\iff\quad \frakM \models \varphi\,.
\end{align*}

To make this proof work for finite trees, it is sufficient to modify the construction
of the system~$\widehat\frakM$.
A closer look at the proof of Lemma~12 reveals that it does not require
infinite branching for~$\widehat\frakM$.
It is enough if we duplicate each subtree sufficiently often,
where the exact number of copies only depends on the formula~$\varphi$.
(Note that there is a remark after Corollary~14 of~\cite{JaninWalukiewicz96}
indicating that Janin and Walukiewicz were already aware of this fact.)
\end{proof}

As a preparation for the more involved characterisation results to follow,
we simplify our task by introducing the following property of a class~$\calC$
of transition systems, which will turn out to be equivalent to having a
characterisation result for bisim\-u\-la\-tion-invari\-ant $\MSO$ over~$\calC$.
\begin{Def}
We say that a class~$\calC$ of transition systems has the
\emph{unravelling property} if, for every $\MSO$-formula~$\varphi$ that is
bisim\-u\-la\-tion-invari\-ant over~$\calC$, there exists an
$\MSO$-formula~$\hat\varphi$ that is bisim\-u\-la\-tion-invari\-ant over trees
such that
\begin{align*}
  \frakS \models \varphi \quad\iff\quad \calU(\frakS) \models \hat\varphi\,,
  \quad\text{for all } \frakS \in \calC\,.
\end{align*}
\upqed\markenddef
\end{Def}

Using Theorem~\ref{Thm: bisim MSO over trees}, we can reformulate this definition
as follows. This version will be our main tool to prove characterisation results for
bisim\-u\-la\-tion-invari\-ant $\MSO$\?: it is sufficient to prove that
the given class has the unravelling property.
\begin{Thm}\label{Thm: unvarelling property}
A class~$\calC$ of transition systems has the unravelling property if,
and only if,
over $\calC$ bisim\-u\-la\-tion-invari\-ant $\MSO$ coincides with~$\Lmu$.
\end{Thm}
\begin{proof}
$(\Rightarrow)$
Suppose that $\calC$~has the unravelling property
and let $\varphi \in \MSO$ be bisim\-u\-la\-tion-invari\-ant over~$\calC$.
Then there exists an $\MSO$-formula $\hat\varphi$ that is
bisim\-u\-la\-tion-invari\-ant over trees and satisfies
\begin{align*}
  \frakS \models \varphi \quad\iff\quad \calU(\frakS) \models \hat\varphi\,,
  \quad\text{for all } \frakS \in \calC\,.
\end{align*}
We can use Theorem~\ref{Thm: bisim-inv MSO over trees}
to find an $\Lmu$-formula~$\psi$ such that
\begin{align*}
  \frakT \models \hat\varphi \quad\iff\quad \frakT \models \psi\,,
  \quad\text{for all trees } \frakT\,.
\end{align*}
For $\frakS \in \calC$, it follows by bisim\-u\-la\-tion-invari\-ance of~$\Lmu$ that
\begin{align*}
  \frakS \models \varphi
  \quad\iff\quad \calU(\frakS) \models \hat\varphi
  \quad\iff\quad \calU(\frakS) \models \psi
  \quad\iff\quad \frakS \models \psi\,.
\end{align*}

$(\Leftarrow)$
Suppose that, over $\calC$, bisim\-u\-la\-tion-invari\-ant $\MSO$ coincides
with~$\Lmu$.
To show that $\calC$~has the unravelling property, consider an
$\MSO$-formula~$\varphi$ that is bisim\-u\-la\-tion-invari\-ant over~$\calC$.
By assumption, there exists an $\Lmu$-formula~$\psi$ such that
\begin{align*}
  \frakS \models \varphi \quad\iff\quad \frakS \models \psi\,,
  \quad\text{for } \frakS \in \calC\,.
\end{align*}
Let $\hat\varphi$ be an $\MSO$-formula that is equivalent to~$\psi$ over every
transition system.
As $\psi$~is bisim\-u\-la\-tion-invari\-ant over all transition systems,
the formula~$\hat\varphi$ is bisim\-u\-la\-tion-invari\-ant over trees and
we have
\begin{align*}
  \frakS \models \varphi
  \quad\iff\quad \frakS \models \psi
  \quad\iff\quad \calU(\frakS) \models \psi
  \quad\iff\quad \calU(\frakS) \models \hat\varphi\,,
\qquad\text{for all } \frakS \in \calC\,.
\end{align*}
\upqed
\end{proof}

Let us also note the following result, which allows us to extend the unravelling
property from a given class to certain superclasses.
\begin{Lem}\label{Lem: extending the unravelling property}
Let $\calC_0 \subseteq \calC$ be classes such that every system in~$\calC$
is bisimilar to one in~$\calC_0$.
If $\calC_0$~has the unravelling property, then so does~$\calC$.
\end{Lem}
\begin{proof}
Let $\varphi$~be bisim\-u\-la\-tion-invari\-ant over~$\calC$.
Then it is also bisim\-u\-la\-tion-invari\-ant over~$\calC_0$ and we can find a
formula~$\hat\varphi$ that is bisim\-u\-la\-tion-invari\-ant over trees such
that
\begin{align*}
  \frakS \models \varphi \quad\iff\quad \calU(\frakS) \models \hat\varphi\,,
  \quad\text{for all } \frakS \in \calC_0\,.
\end{align*}
We claim that this formula has the desired properties.
Thus, consider a system $\frakS \in \calC$.
By assumption, we have $\frakS \bisim \frakS_0$ for some $\frakS_0 \in \calC_0$.
By Lemma~\ref{Lem: unravelling preserves bisimulation}, it follows that
$\calU(\frakS) \bisim \calU(\frakS_0)$.
Consequently, by bisim\-u\-la\-tion-invari\-ance of $\varphi$~over~$\calC$
and of $\hat\varphi$~over trees, we have
\begin{align*}
  \frakS \models \varphi
  \quad\iff\quad \frakS_0 \models \varphi
  \quad\iff\quad \calU(\frakS_0) \models \hat\varphi
  \quad\iff\quad \calU(\frakS) \models \hat\varphi\,.
\end{align*}
\upqed
\end{proof}

\section{Composition lemmas}   %%%%%%%%%%%%%%%%%%%%%%%%%%%%%%%%%%%%%%%%%%%%%%%%%%%%%%%%%%%
\label{Sect: composition}

We have mentioned above that automata-theoretic methods have so far been unsuccessful
at attacking the finite version of the Janin--Walukiewicz result.
Therefore, we rely on the composition method instead.
Let us recall how this method works.

\begin{Def}
Let $\frakS$~and~$\frakT$~be transition systems (or general structures) and
$m < \omega$ a number.
The \emph{$m$-theory} $\Th_m(\frakS)$ of~$\frakS$ is the set of all
$\MSO$-formulae of quantifier-rank~$m$ that are satisfied by~$\frakS$.
(The quantifier-rank of a formula is its nesting depths of (first-order and second-order)
quantifiers.)
We write
\begin{align*}
  \frakS \equiv_m \frakT \quad\defiff\quad \Th_m(\frakS) = \Th_m(\frakT)\,.
\end{align*}
\upqed
\markenddef
\end{Def}

Roughly speaking the composition method provides some machinery that allows us
to compute the $m$-theory of a given transition system
by breaking it down into several components
and looking at the $m$-theories of these components separately.
This approach is based on the realisation that
several operations on transition systems are compatible with
$m$-theories in the sense that the $m$-theory of the result
can be computed from the $m$-theories of the arguments.
Statements to that effect are known as \emph{composition theorems.}
For an overview we refer the reader to \cite{BlumensathColcombetLoeding07}
and~\cite{Makowsky04}.
The following basic operations and their composition theorems will be used below.
We start with disjoint unions.
\begin{Def}
The \emph{disjoint union} of two structures
$\frakA = \langle A,R_0^\frakA,\dots,R_m^\frakA\rangle$ and
$\frakB = \langle B,R_0^\frakB,\dots,R_m^\frakB\rangle$
is the structure
\begin{align*}
  \frakA \oplus \frakB :=
    \bigl\langle A \cupdot B,\ R_0^\frakA \cupdot R_0^\frakB,\dots,\
                         R_m^\frakA \cupdot R_m^\frakB,\
           \mathrm{Left},\ \mathrm{Right}\bigr\rangle
\end{align*}
obtained by forming the disjoint union of the universes and relations
of $\frakA$~and~$\frakB$ and adding two unary predicates
$\mathrm{Left} := A$
and
$\mathrm{Right} := B$
that mark whether an element belongs to~$\frakA$ or to~$\frakB$.
If $\frakA$~and~$\frakB$ are transition systems, the initial state of
$\frakA \oplus \frakB$ is that of~$\frakA$.
\markenddef
\end{Def}
The corresponding composition theorem looks as follows.
It can be proved by a simple induction on~$m$.
\begin{Lem}\label{Lem: composition for disjoint union}
$\frakA \equiv_m \frakA' \qtextq{and} \frakB \equiv_m \frakB'
 \qtextq{implies}
 \frakA \oplus \frakB \equiv_m \frakA' \oplus \frakB'\,.$
\end{Lem}

Two other operations we need are interpretations and fusion operations.
\begin{Def}
An \emph{interpretation} is an operation~$\tau$ on structures that is given by
a list $\langle\delta(x),(\varphi_R(\bar x))_{R \in \Sigma}\rangle$
of $\MSO$-formulae. Given a structure~$\frakA$, it produces the structure
$\tau(\frakA)$ whose universe consists of all elements of~$\frakA$ satisfying
the formula~$\delta$ and whose relations are those defined by the
formulae~$\varphi_R$.
The \emph{quantifier-rank} of an interpretation is the maximal quantifier-rank
of a formula in the list.
An interpretation is \emph{quantifier-free} if its quantifier-rank is~$0$.
\markenddef
\end{Def}
\begin{Lem}
Let $\tau$~be an interpretation of quantifier-rank~$k$. Then
\begin{align*}
  \frakA \equiv_{m+k} \frakA' \qtextq{implies}
  \tau(\frakA) \equiv_m \tau(\frakA')\,.
\end{align*}
\end{Lem}

\begin{Def}
Let $P$~be a predicate symbol.
The \emph{fusion operation} $\fuse_P$ merges in a given structure
all elements of the set~$P$ into a single element,
i.e., all elements of~$P$ are replaced by a single new element and all
edges incident with one of the old elements are attached to the new one
instead.
\markenddef
\end{Def}
\begin{Lem}
$\frakA \equiv_m \frakA' \qtextq{implies} \fuse_P(\frakA) \equiv_m \fuse_P(\frakA')\,.$
\end{Lem}

Using the composition theorems for these basic operations we can prove new
theorems for derived operations.
As an example let us consider \emph{pointed paths,}
i.e., paths where both end-points are marked by special colours.
\begin{Def}
We denote the \emph{concatenation} of two paths $\frakA$~and~$\frakB$
by $\frakA + \frakB$.
And we write $\frakA^\bullet$ for the expansion of a path~$\frakA$ by two new constants
for the end-points.
\markenddef
\end{Def}
\begin{Cor}\label{Cor: composition for concatenation}
Let $\frakA,\frakA',\frakB,\frakB'$ be paths.
Then
$\frakA^\bullet \equiv_m \frakA'^\bullet$ and $\frakB^\bullet \equiv_m \frakB'^\bullet$
implies
$(\frakA + \frakB)^\bullet \equiv_m (\frakA' + \frakB')^\bullet\,.$
\end{Cor}
\begin{proof}
As the end-points are given by constants,
we can construct a quantifier-free interpretation~$\tau$ mapping
$\frakA^\bullet \oplus \frakB^\bullet$ to $(\frakA + \frakB)^\bullet$.
\end{proof}
Note that, since the concatenation operation is associative, it in particular
follows that the set of $m$-theories of paths forms a semigroup.

Finally let us mention one more involved operation with a composition theorem.
Let $\frakS$~be a transition system and $\frakC \subseteq \frakS$ a subsystem.
We say that $\frakC$~is \emph{attached} at the state $s \in S$ if
there is a unique edge (in either direction) between a state in $S \setminus C$
and a state in~$C$ and this edge leads from~$s$ to the initial state of~$\frakC$.
\begin{Prop}\label{Prop: generalised sum}
Let $\frakS$~be a (possibly infinite) transition system
and let $\frakS'$~be the system obtained from~$\frakS$ by replacing
an arbitrary number of attached subsystems by subsystems with the same $m$-theories
(as the corresponding replaced ones).
Then $\frakS \equiv_m \frakS'$.
\end{Prop}
For a finite system~$\frakS$ this statement can be proved in the same way
as Corollary~\ref{Cor: composition for concatenation} by expressing~$\frakS$
as a disjoint union followed by a quantifier-free interpretation.
For infinite systems, we need a more powerful version of the disjoint union operation
called a \emph{generalised sum} (see~\cite{Shelah75}).

As presented above these tools work with $m$-theories, which is not quite what we need
since we have to also account for bisim\-u\-la\-tion-invari\-ance.
To do so we modify the definitions as follows.
\begin{Def}
Let $\calC$~be a class of transition systems and $m < \omega$ a number.

(a) We denote by~$\simeq^m_\calC$ the transitive closure of the union
${\equiv_m} \cup {\bisim}$ restricted to the class~$\calC$.
Formally, we define
$\frakS \simeq^m_\calC \frakT$
if there exist systems $\frakC_0,\dots,\frakC_n \in \calC$ such that
\begin{align*}
  \frakC_0 = \frakS\,,\quad
  \frakC_n = \frakT\,,
  \qtextq{and}
  \frakC_i \equiv_m \frakC_{i+1}
  \qtextq{or}
  \frakC_i \bisim \frakC_{i+1}\,,
  \qquad\text{for all } i < n\,.
\end{align*}

(b) We denote by $\Th^m_\calC(\frakS)$ the set of all $\MSO$-formulae
of quantifier-rank~$m$ that are bisim\-u\-la\-tion-invari\-ant over~$\calC$
and that are satisfied by~$\frakS$, and we define
\begin{align*}
  \frakS \equiv^m_\calC \frakS'
  \quad\defiff\quad
  \Th^m_\calC(\frakS) = \Th^m_\calC(\frakS')\,.
\end{align*}
We also set
$\TH^m_\calC := \set{ \Th^m_\calC(\frakS) }{ \frakS \in \calC }\,.$
\markenddef
\end{Def}

Note that, up to logical equivalence, there are only finitely many formulae of a given
quantifier-rank. Hence, each set $\TH^m_\calC$ is finite and the relations
$\equiv_m$,~$\equiv^m_\calC$ and $\simeq^m_\calC$ have finite index.

The relation~$\equiv^m_\calC$ is what we aim to understand when proving
characterisation results. But there is no obvious way to compute it.
As an approximation we have introduced the relation~$\simeq^m_\calC$,
which is defined in terms of relations that we hopefully understand much better.
Surprisingly, our approximation turns out to be exact.
\begin{Prop}\label{Prop: simeq and bisim. theories}
The relations $\simeq^m_\calC$ and $\equiv^m_\calC$ coincide.
\end{Prop}
\begin{proof}
Clearly $\frakS \simeq^m_\calC \frakT$ implies $\frakS \equiv^m_\calC \frakT$
as no bisim\-u\-la\-tion-invari\-ant $\MSO$-formula of quantifier rank at most~$m$
can distinguish two $\simeq^m_\calC$-equivalent transition systems.
To prove the converse we consider the formulae
\begin{align*}
  \psi_\frakC := \Lor {\bigset{\textstyle \Land \Th_m(\frakS) }{ \frakC \simeq^m_\calC \frakS }}\,,
  \qquad\text{for }\frakC \in \calC\,.
\end{align*}
(This is well-defined since, up to logical equivalence, there are only finitely many
$m$-theories and each of them only contains finitely many formulae.)
We start by showing that
\begin{align*}
  \frakT \models \psi_\frakC \quad\iff\quad \frakT \simeq^m_\calC \frakC\,.
\end{align*}
Clearly, $\frakT \simeq^m_\calC \frakC$ implies $\frakT \models \psi_\frakC$
by definition of~$\psi_\frakC$.
Conversely,
\begin{align*}
  \frakT \models \psi_\frakC
  &\quad\Rightarrow\quad
  \frakT \models \Th_m(\frakS) \text{ for some } \frakS \text{ with } \frakS \simeq^m_\calC \frakC \\
  &\quad\Rightarrow\quad
  \frakT \equiv_m \frakS \text{ for some } \frakS \text{ with } \frakS \simeq^m_\calC \frakC \\
  &\quad\Rightarrow\quad
  \frakT \simeq^m_\calC \frakC\,.
\end{align*}

Furthermore, note that $\psi_\frakC$ is bisimulation-invariant over~$\calC$ since
\begin{align*}
  \frakS \sim \frakT
  \quad\Rightarrow\quad
  \frakS \simeq^m_\calC \frakT
  \quad\Rightarrow\quad
  (\frakS \models \psi_\frakC \Leftrightarrow \frakT \models \psi_\frakC)\,.
\end{align*}
Thus, $\psi_\frakC$ is an $\MSO_m$-formula that is bisimulation-invariant over~$\calC$,
and it follows that
\begin{align*}
  \frakS \equiv^m_\calC \frakT
  &\quad\Rightarrow\quad
  (\forall\frakC \in \calC)[\frakS \models \psi_\frakC \Leftrightarrow \frakT \models \psi_\frakC] \\
  &\quad\Rightarrow\quad
  \frakT \models \psi_\frakS \\
  &\quad\Rightarrow\quad
  \frakS \simeq^m_\calC \frakT\,.
\end{align*}
\upqed
\end{proof}

Some of the above composition theorems also hold for the relation~$\simeq^m_\calC$.
This is immediate if the operation in question also preserves bisimilarity.
We mention only two such results. The second one will be needed below.
\begin{Lem}
Let $\calC$~be a class that is closed under disjoint unions.
\begin{align*}
  \frakA \simeq^m_\calC \frakA' \qtextq{and} \frakB \simeq^m_\calC \frakB'
  \qtextq{implies}
  \frakA \oplus \frakB \simeq^m_\calC \frakA' \oplus \frakB'\,.
\end{align*}
\end{Lem}
\begin{Prop}\label{Prop: replacing attached subsystems}
Let $\calC$~and~$\calD$ be two classes,
$\frakS \in \calC$~a (possibly infinite) transition system
and let $\frakS'$~be the system obtained from~$\frakS$ by replacing
an arbitrary number of attached subsystems by subsystems which are
$\simeq^m_\calD$-equivalent.
Then $\frakS \simeq^m_\calC \frakS'$ provided that the class~$\calC$
is closed under the operation of replacing attached subsystems in~$\calD$.
\end{Prop}

\section{Types}   %%%%%%%%%%%%%%%%%%%%%%%%%%%%%%%%%%%%%%%%%%%%%%%%%%%%%%%%%%%%%%%%%%%%%%%%%%%%%%%%%%
\label{Sect: types}

Our strategy to prove the unravelling property for a class~$\calC$ is as follows.
For every quan\-ti\-fier-rank~$m$, we assign to each tree~$\frakT$
a so-called \emph{$m$-type}~$\tau_m(\frakT)$. We choose the functions~$\tau_m$ such that
we can compute the theory $\Th^m_\calC(\frakC)$ of a system $\frakC \in\nobreak \calC$
from the $m$-type $\tau_m(\calU(\frakC))$ of its unravelling.
Furthermore, we need to find $\MSO$-formulae checking whether a tree has a given $m$-type.
The formal definition is as follows.
\begin{Def}
Let $\calC$~be a class of transition systems and $\calT$~the class of all trees.

(a)
A \emph{family of type functions} for~$\calC$ is a family of functions
$\tau_m : \calT \to \Theta_m$, for $m < \omega$, where the co-domains~$\Theta_m$
are finite sets and each $\tau_m$~satisfies the following two axioms.
\begin{enumerate}[label={\normalfont\scshape(S\arabic*)}]
\item $\tau_m(\calU(\frakC)) = \tau_m(\calU(\frakC'))
  \!\qtextq{implies}\!
  \Th^m_\calC(\frakC) = \Th^m_\calC(\frakC')\,,
  \quad\text{for } \frakC,\frakC' \in\nobreak \calC\,.$
\item $\frakT \bisim \frakT' \qtextq{implies}
       \tau_m(\frakT) = \tau_m(\frakT')\,,
       \quad\text{for all } \frakT,\frakT' \in \calT\,.$
\end{enumerate}

(b) A family $(\tau_m)_m$ of type functions is \emph{definable} if, for every
$\theta \in \Theta_m$, there exists an $\MSO$-formula~$\psi_\theta$ such that
\begin{enumerate}[label={\normalfont\scshape(S\arabic*)}, start=3]
\item $\frakT \models \psi_\theta \quad\iff\quad \tau_m(\frakT) = \theta\,,
  \quad\text{for all trees } \frakT\,.$
\end{enumerate}
\upqed
\markenddef
\end{Def}

Let us start by showing how to prove the unravelling property using type functions.
The following characterisation theorem can be considered to be the main theoretical result
of this article.
\begin{Thm}\label{Thm: type functions imply unravelling}
Let $\calC$~be a class of transition systems and $\calT$~the class of all trees.
The following statements are equivalent.
\begin{enum1}
\item Over~$\calC$, bisim\-u\-la\-tion-invari\-ant\/ $\MSO$ coincides with\/~$\Lmu$.
\item $\calC$ has the unravelling property.
\item There exists a definable family $(\tau_m)_m$ of type functions for~$\calC$.
\item The $g(m)$-theory of~$\calU(\frakC)$ determines the $m$-theory of~$\frakC$ in the sense
  that there exist functions $g : \omega \to \omega$ and
  $h_m : \TH^{g(m)}_\calT \to \TH^m_\calC$, for $m < \omega$, such that
  \begin{align*}
    h_m\bigl(\Th^{g(m)}_\calT(\calU(\frakC))\bigr) = \Th^m_\calC(\frakC)\,,
    \quad\text{for all } \frakC \in \calC\,.
  \end{align*}
\end{enum1}
\end{Thm}
\begin{proof}
(1)~$\Leftrightarrow$~(2) was already proved in Theorem~\ref{Thm: unvarelling property}.

(2)~$\Rightarrow$~(4)
Let $m < \omega$. For every $\theta \in \TH^m_\calC$,
we use the unravelling property to find an $\MSO$-formula~$\varphi_\theta$
that is bisim\-u\-la\-tion-invari\-ant over trees and satisfies
\begin{align*}
  \frakC \models \Land\theta
  \quad\iff\quad
  \calU(\frakC) \models \varphi_\theta\,,
  \quad\text{for } \frakC \in \calC\,.
\end{align*}
Let $k$~be the maximal quantifier-rank of these formulae~$\varphi_\theta$.
Then
\begin{align*}
  \Th^k_\calT(\calU(\frakC)) = \Th^k_\calT(\calU(\frakC'))
  \qtextq{implies}
  \Th^m_\calC(\frakC) = \Th^m_\calC(\frakC')\,.
\end{align*}
Consequently, there exists a function $h_m : \TH^k_\calT \to \TH^m_\calC$
such that
\begin{align*}
  h_m\bigl(\Th^k_\calT(\calU(\frakC))\bigr) = \Th^m_\calC(\frakC)\,.
\end{align*}

(4)~$\Rightarrow$~(3)
Given $h_m : \TH^k_\calT \to \TH^m_\calC$, we set
\begin{align*}
  \tau_m(\frakT) := h_m\bigl(\Th^{g(m)}_\calT(\frakT)\bigr)\,.
\end{align*}
We claim that $(\tau_m)_m$~is a definable family of type functions.
For \textsc{(S1),} suppose that $\tau_m(\calU(\frakC)) = \tau_m(\calU(\frakC'))$.
Then
\begin{align*}
  \Th^m_\calC(\frakC)
   = h_m\bigl(\Th^{g(m)}_\calT(\calU(\frakC))\bigr)
   = h_m\bigl(\Th^{g(m)}_\calT(\calU(\frakC'))\bigr)
   = \Th^m_\calC(\frakC')\,.
\end{align*}
For \textsc{(S2),} suppose that $\frakT \bisim \frakT'$. Then
\begin{align*}
  \Th^{g(m)}_\calT(\frakT) = \Th^{g(m)}_\calT(\frakT')\,,
  \qtextq{which implies that}
  \tau_m(\frakT) = \tau_m(\frakT')\,.
\end{align*}
For \textsc{(S3),} set
\begin{align*}
  \psi_\theta := \Lor {\bigset{ \textstyle\Land\Delta }{ \Delta \in h_m^{-1}(\theta) }}\,,
  \quad\text{for } \theta \in \TH^m_\calC\,.
\end{align*}
Then
\begin{align*}
  \frakT \models \psi_\theta
  \quad\iff\quad
   \Th^{g(m)}_\calT(\frakT) \in h_m^{-1}(\theta)
  \quad\iff\quad
   h_m\bigl(\Th^{g(m)}_\calT(\frakT)\bigr) = \theta
   \quad\iff\quad
   \tau_m(\frakT) = \theta\,.
\end{align*}

(3)~$\Rightarrow$~(4)
Let $\psi_\theta$, for $\theta \in \Theta_m$, be the formulae given by \textsc{(S3).}
For each $m < \omega$, let $g(m)$~be the maximal quantifier-rank of~$\psi_\theta$,
for $\theta \in \Theta_m$.

We start by showing that each~$\psi_\theta$ is bisim\-u\-la\-tion-invari\-ant over
trees\?: given $\frakT \bisim \frakT'$, \textsc{(S2)} implies that
\begin{align*}
  \frakT \models \psi_\theta
  \quad\iff\quad
  \tau_m(\frakT) = \theta
  \quad\iff\quad
  \tau_m(\frakT') = \theta
  \quad\iff\quad
  \frakT' \models \psi_\theta\,,
\end{align*}
as desired.
By the claim we have just proved, it follows that
\begin{align*}
  \frakT \equiv^{g(m)}_\calT \frakT'
  \qtextq{implies}
  \tau_m(\frakT) = \tau_m(\frakT')\,.
\end{align*}
Consequently, there exist functions $f_m : \TH^{g(m)}_\calT \to \Theta_m$ such that
\begin{align*}
  f_m\bigl(\Th^{g(m)}_\calT(\calU(\frakC))\bigr)
  = \tau_m(\calU(\frakC))\,.
\end{align*}
By \textsc{(S1),} we can find functions $\sigma_m : \Theta_m \to \TH^m_\calC$ such that
\begin{align*}
  \sigma_m(\tau_m(\calU(\frakC))) = \Th^m_\calC(\frakC)\,.
\end{align*}
Setting $h_m := \sigma_m \circ f_m$ it follows that
\begin{align*}
  h_m\bigl(\Th^{g(m)}_\calT(\calU(\frakC))\bigr)
  = \sigma_m\bigl(f_m\bigl(\Th^{g(m)}_\calT(\calU(\frakC))\bigr)\bigr)
  = \sigma_m\bigl(\tau_m(\calU(\frakC))\bigr)
  = \Th^m_\calC(\frakC)\,.
\end{align*}

(4)~$\Rightarrow$~(2)
Let $\varphi$~be an $\MSO$-formula of quantifier-rank~$m$ that is
bisim\-u\-la\-tion-invari\-ant over~$\calC$.
We claim that the formula
\begin{align*}
  \hat\varphi :=
    \Lor {\bigset{ \textstyle\Land\theta }
                 { \theta \in \TH^{g(m)}_\calT,\ \varphi \in h_m^{-1}(\theta) }}
\end{align*}
has the desired properties.
First of all,
\begin{align*}
  \calU(\frakC) \models \hat\varphi
  &\quad\iff\quad
  \Th^{g(m)}_\calT(\calU(\frakC)) = \theta \text{ for some } \theta
  \text{ with } \varphi \in h_m(\theta) \\
  &\quad\iff\quad
   \varphi \in h_m\bigl(\Th^{g(m)}_\calT(\calU(\frakC))\bigr) = \Th^m_\calC(\frakC) \\
  &\quad\iff\quad
   \frakC \models \varphi\,.
\end{align*}
Hence, it remains to show that $\hat\varphi$~is bisim\-u\-la\-tion-invari\-ant over
trees. Let $\frakT \bisim \frakT'$. Then
$\Th^{g(m)}_\calT(\frakT) = \Th^{g(m)}_\calT(\frakT')$ and we have
\begin{align*}
  \frakT \models \hat\varphi
  \quad\iff\quad
   \varphi \in h_m\bigl(\Th^{g(m)}_\calT(\frakT)\bigr)
  \quad\iff\quad
   \varphi \in h_m\bigl(\Th^{g(m)}_\calT(\frakT')\bigr)
   \quad\iff\quad
   \frakT' \models \hat\varphi\,.
\end{align*}
\upqed
\end{proof}

\section{Lassos}   %%%%%%%%%%%%%%%%%%%%%%%%%%%%%%%%%%%%%%%%%%%%%%%%%%%%%%%%%%%%%%%%%%%%%%%%%%%%%%%%%
\label{Sect: lassos}

As an application of type functions, we consider a very simple example,
the class of \emph{lassos.} Our proof is based on more or less the same arguments
as that by Dawar and Janin~\cite{DawarJanin04}, just the presentation differs.
A lasso is a transition system consisting of a directed path ending in a cycle.
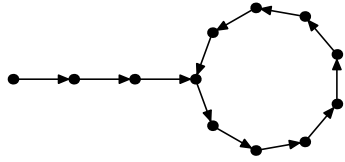
\begin{center}
\includegraphics{Bisimulation-submitted-1.mps}
%\begin{mpfig}
%  u := 0.8cm;
%
%  z0  = (0,0);
%  z1  = (u,0);
%  z2  = (2u,0);
%  z3  = (3u,0);
%  z4  = z3 + 1.2u*right;
%  z5  = z4 + 1.2u*dir 220;
%  z6  = z4 + 1.2u*dir 260;
%  z7  = z4 + 1.2u*dir 300;
%  z8  = z4 + 1.2u*dir 340;
%  z9  = z4 + 1.2u*dir  20;
%  z10 = z4 + 1.2u*dir  60;
%  z11 = z4 + 1.2u*dir 100;
%  z12 = z4 + 1.2u*dir 140;
%
%  pickup pencircle scaled 0.6pt;
%
%  drawarrow anchor(z0,z1,2pt)   -- anchor(z1,z0,2pt);
%  drawarrow anchor(z1,z2,2pt)   -- anchor(z2,z1,2pt);
%  drawarrow anchor(z2,z3,2pt)   -- anchor(z3,z2,2pt);
%  drawarrow anchor(z3,z5,2pt)   -- anchor(z5,z3,2pt);
%  drawarrow anchor(z5,z6,2pt)   -- anchor(z6,z5,2pt);
%  drawarrow anchor(z6,z7,2pt)   -- anchor(z7,z6,2pt);
%  drawarrow anchor(z7,z8,2pt)   -- anchor(z8,z7,2pt);
%  drawarrow anchor(z8,z9,2pt)   -- anchor(z9,z8,2pt);
%  drawarrow anchor(z9,z10,2pt)  -- anchor(z10,z9,2pt);
%  drawarrow anchor(z10,z11,2pt) -- anchor(z11,z10,2pt);
%  drawarrow anchor(z11,z12,2pt) -- anchor(z12,z11,2pt);
%  drawarrow anchor(z12,z3,2pt)  -- anchor(z3,z12,2pt);
%
%  pickup pencircle scaled 4pt;
%
%  draw z0;
%  draw z1;
%  draw z2;
%  draw z3;
%  draw z5;
%  draw z6;
%  draw z7;
%  draw z8;
%  draw z9;
%  draw z10;
%  draw z11;
%  draw z12;
%\end{mpfig}
\end{center}
We allow the borderline cases where the initial path has length~$0$ or
the cycle consists of only a single edge.

To define the type of a lasso, note that we can construct every lasso~$\frakL$
from two finite paths $\frakA$~and~$\frakB$ by identifying three of their
end-points.
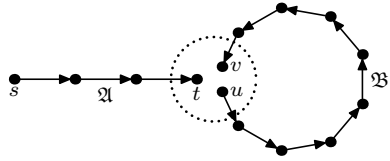
\begin{center}
\includegraphics{Bisimulation-submitted-2.mps}
\end{center}
The paths $\frakA$~and~$\frakB$ are uniquely determined by~$\frakL$.
We will refer to $\frakA$~as the \emph{tail} of the lasso and to~$\frakB$
as the \emph{loop.}
We introduce two kinds of types for lassos, a strong one and a weak one.
\begin{Def}
The \emph{strong $m$-type} of a lasso~$\frakL$ with tail $\frakA$ and loop~$\frakB$
is the pair
\begin{align*}
  \stp_m(\frakL) := \langle\alpha,\beta\rangle\,,
  \quad\text{where}\quad
  \alpha := \Th_m(\frakA^\bullet)
  \quad\text{and}\quad
  \beta := \Th_m(\frakB^\bullet)\,.
\end{align*}
\upqed
\markenddef
\end{Def}
The strong $m$-type of a lasso uniquely determines its $m$-theory.
\begin{Lem}\label{Lem: strong type determines m-theory}
Let $\frakL_0$~and~$\frakL_1$ be lassos.
\begin{align*}
  \stp_m(\frakL_0) = \stp_m(\frakL_1)
  \qtextq{implies}
  \frakL_0 \equiv_m \frakL_1\,.
\end{align*}
\end{Lem}
\begin{proof}
Let $\frakA_i$~and~$\frakB_i$ be the tail and loop of~$\frakL_i$.
Note that we can write~$\frakL_i$ in the form
\begin{align*}
  \frakL_i = \fuse_{P_i}\bigl(\langle\frakA_i,s_it_i,P_i\rangle \oplus
                              \langle\frakB_i,u_iv_i,P_i\rangle\bigr)\,,
\end{align*}
where $s_i,t_i,u_i,v_i$ are the respective end-points of $\frakA_i$~and~$\frakB_i$,
$P_i = \{t_i,u_i,v_i\}$ is an additional unary predicate marking the vertices to be
identified, and $\fuse_{P_i}$ is the \emph{fusion operation} that identifies all
vertices in~$P_i$.
Note that $P_i$~is definable by a quan\-ti\-fier-free formula.
Hence, there exists a quan\-ti\-fier-free interpretation~$\sigma$ such that
\begin{align*}
  \frakL_i = \fuse_{P_i}\bigl(\sigma\bigl(\langle\frakA_i^\bullet\rangle \oplus
                                          \langle\frakB_i^\bullet\rangle\bigr)\bigr)\,.
\end{align*}
As disjoint union, quan\-ti\-fier-free interpretations, and fusion are
compatible with $m$-theories, it follows that
$\frakA_0^\bullet \equiv_m \frakA_1^\bullet$ and
$\frakB_0^\bullet \equiv_m \frakB_1^\bullet$ implies
\begin{align*}
  &\frakL_0
  = \fuse_{P_0}\bigl(\sigma\bigl(\frakA_0^\bullet \oplus
                                 \frakB_0^\bullet\bigr)\bigr)
  \equiv_m
  \fuse_{P_1}\bigl(\sigma\bigl(\frakA_1^\bullet \oplus
                               \frakB_1^\bullet\bigr)\bigr)
   = \frakL_1\,.
\end{align*}
\upqed
\end{proof}

The problem with the strong type of a lasso~$\frakL$ is that we cannot recover it
from the unravelling of~$\frakL$ as the decomposition of $\calU(\frakL)$ into the
parts of~$\frakL$ is uncertain.
Therefore we introduce another notion of a type where this recovery is possible.
For this we recall some facts from the theory of $\omega$-semigroups.

Recall that we have noted in Corollary~\ref{Cor: composition for concatenation}
that the $m$-theories of pointed paths form a finite semigroup
with respect to concatenation. Furthermore, every element~$a$
of a finite semigroup has an \emph{idempotent power}~$a^\pi$, which is defined
as the value~$a^n$ where $n$~is the least natural number such that
$a^n \cdot a^n = a^n$.
\begin{Def}
(a) A \emph{factorisation} of an infinite path~$\frakA$
is a sequence $(\frakA_i)_{i<\omega}$ of finite paths whose concatenation is~$\frakA$.
Such a factorisation has \emph{$m$-type} $\langle\alpha,\beta\rangle$ if
\begin{align*}
  \alpha := \Th_m(\frakA_0^\bullet)
  \qtextq{and}
  \beta := \Th_m(\frakA_i^\bullet)\,, \quad\text{for } i > 0\,.
\end{align*}

(b) Two pairs $\langle\alpha,\beta\rangle$ and
$\langle\gamma,\delta\rangle$ of $m$-theories are \emph{conjugate}
if there are $m$-theories $\xi$~and~$\eta$ such that
\begin{align*}
  \gamma\delta^\pi = \alpha\beta^\pi\xi\,,\quad
  \beta^\pi = \xi\eta\,,
  \qtextq{and}
  \delta^\pi = \eta\xi\,.
\end{align*}
Being conjugate is an equivalence relation.
We denote the equivalence class of a pair
$\langle\alpha,\beta\rangle$ by $[\alpha,\beta]$.

(c)
The \emph{weak $m$-type} of a lasso~$\frakL$ with parts $\frakA$~and~$\frakB$
is
\begin{align*}
  \wtp_m(\frakL) := [\alpha,\beta]\,,
  \quad\text{where}\quad
  \alpha := \Th_m(\frakA^\bullet)
  \quad\text{and}\quad
  \beta := \Th_m(\frakB^\bullet)\,.
\end{align*}

(d) The \emph{$m$-type} of an infinite tree~$\frakT$ is
\begin{align*}
  \tau_m(\frakT) := [\alpha,\beta]\,,
\end{align*}
where $\alpha$~and~$\beta$ is an arbitrary pair of $m$-theories such that
every branch of~$\frakT$ has a factorisation of $m$-type $\langle\alpha,\beta\rangle$.
If there is no such pair, we set $\tau_m(\frakT) := \bot$.
\markenddef
\end{Def}

\begin{Lem}\label{Lem: lasso type determines m-L-equivalence}
Let\/ $\calL$~be the class of all lassos and
let\/ $\frakL_0, \frakL_1 \in \calL$.
\begin{align*}
  \wtp_m(\frakL_0) = \wtp_m(\frakL_1)
  \qtextq{implies}
  \frakL_0 \simeq^m_\calL \frakL_1\,.
\end{align*}
\end{Lem}
\begin{proof}
Let $\frakA_i$~and~$\frakB_i$ be the parts of the lasso~$\frakL_i$,
and set
\begin{align*}
  \alpha_i := \Th_m(\frakA_i^\bullet)
  \qtextq{and}
  \beta_i := \Th_m(\frakB_i^\bullet)\,.
\end{align*}
Since the pairs $\langle\alpha_0,\beta_0\rangle$
and $\langle\alpha_1,\beta_1\rangle$ are conjugate, there exist
$m$-theories $\xi$~and~$\eta$ such that
\begin{align*}
  \alpha_1\beta_1^\pi = \alpha_0\beta_0^\pi\xi\,,
  \quad
  \beta_0^\pi = \xi\eta\,,
  \qtextq{and}
  \beta_1^\pi = \eta\xi\,.
\end{align*}
Fix exponents $k_0$~and~$k_1$ such that $\beta_i^\pi = \beta_i^{k_i}$
and let $\frakC$~and~$\frakD$ be finite paths with
\begin{align*}
  \xi = \Th_m(\frakC^\bullet)
  \qtextq{and}
  \eta = \Th_m(\frakD^\bullet)\,.
\end{align*}
We construct lassos $\frakM_0$,~$\frakM_1$, $\frakN_0$, and~$\frakN_1$
as follows.
The lasso~$\frakM_i$ has the parts
\begin{align*}
  \frakA_i + \frakB_i^{k_i}
  \qtextq{and}
  \frakB_i^{k_i}\,,
\end{align*}
$\frakN_0$~has the parts
\begin{align*}
  \frakA_0 + \frakB_0^{k_0}
  \qtextq{and}
  \frakC + \frakD\,,
\end{align*}
and $\frakN_1$~has the parts
\begin{align*}
  \frakA_0 + \frakB_0^{k_0} + \frakC
  \qtextq{and}
  \frakD + \frakC\,.
\end{align*}
Then $\stp_m(\frakM_i) = \stp_m(\frakN_i)$ and
it follows by Lemma~\ref{Lem: strong type determines m-theory} that
\begin{align*}
  \frakL_0 \bisim \frakM_0
           \equiv_m \frakN_0
           \bisim \frakN_1
           \equiv_m \frakM_1
           \bisim \frakL_1\,.
\end{align*}
\upqed
\end{proof}

To show that the functions $(\tau_m)_m$ form a family of type functions,
we need the following standard facts about factorisations and their types
(see, e.g., Section~II.2 of~\cite{PerrinPin04}).
\begin{Prop}\label{Prop: types of factorisations}
Let\/ $\frakA$~be an infinite path.
\begin{enuma}
\item $\frakA$ has a factorisation of type $\langle\alpha,\beta\rangle$,
  for some $\alpha$~and~$\beta$.
\item If\/ $\frakA$~has factorisations of type $\langle\alpha,\beta\rangle$
  and $\langle\gamma,\delta\rangle$, then
  $\langle\alpha,\beta\rangle$ and $\langle\gamma,\delta\rangle$ are conjugate.
\end{enuma}
\end{Prop}
Note that these two statements imply in particular that the type
$\tau_m(\frakT)$ of a tree~$\frakT$ is well-defined.

\begin{Lem}\label{Lem: tau-m are type family}
The functions $(\tau_m)_m$ defined above form a definable family of type
functions for the class of all lassos.
\end{Lem}
\begin{proof}
\textsc{(S1)}
Suppose that $\tau_m(\calU(\frakL_0)) = \tau_m(\calU(\frakL_1))$, for two
lassos $\frakL_0$~and~$\frakL_1$.
By Proposition~\ref{Prop: types of factorisations}\,(b), it follows that
\begin{align*}
  \wtp_m(\frakL_0)
  = \tau_m(\calU(\frakL_0))
  = \tau_m(\calU(\frakL_1))
  = \wtp_m(\frakL_1)\,.
\end{align*}
Hence, the claim follows by Lemma~\ref{Lem: lasso type determines m-L-equivalence}.

\textsc{(S2)}
Suppose that $\frakT \bisim \frakT'$ and that every branch of~$\frakT$ has
a factorisation of type $\langle\alpha,\beta\rangle$.
Then so does every branch of~$\frakT'$.
Hence, $\tau_m(\frakT) = \tau_m(\frakT')$.

\textsc{(S3)}
Given two $m$-theories $\alpha$~and~$\beta$,
it is straightforward to write down an $\MSO$-formula~$\psi_{\alpha,\beta}$
stating that every branch of a tree has a factorisation of type
$\langle\alpha,\beta\rangle$.
For a conjugacy class $[\alpha,\beta]$, the formula
\begin{align*}
  \varphi_{[\alpha,\beta]} :=
    \Lor_{\langle\gamma,\delta\rangle \in [\alpha,\beta]} \psi_{\alpha,\beta}
\end{align*}
then states that $\tau_m(\frakT) = [\alpha,\beta]$.
\end{proof}

By Theorem~\ref{Thm: type functions imply unravelling},
it therefore follows that the class of lassos has the unravelling property.
\begin{Thm}\label{Thm: characterisation for lassos}
The class of all lassos has the unravelling property.
\end{Thm}

\section{Hierarchical Lassos}   %%%%%%%%%%%%%%%%%%%%%%%%%%%%%%%%%%%%%%%%%%%%%%%%%%%%%%%%%%
\label{Sect: hierarchical lassos}

After the simple example in the previous section, let us give a more substantial
application of the type machinery.
We consider \emph{hierarchical} (or \emph{nested}) lassos.
These are obtained from a lasso by repeatedly attaching sublassos to
some states. More precisely, a $1$-lasso is just an ordinary lasso,
while inductively a $(k+1)$-lasso is obtained from a $k$-lasso by attaching
one or more lassos to some of the states.
(Each state may have several sublassos attached.)
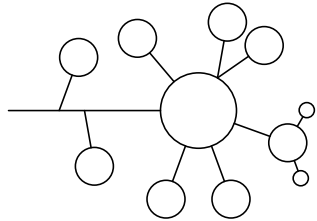
\begin{center}
\includegraphics{Bisimulation-submitted-3.mps}
\end{center}
Alternatively, we can obtain a $(k+1)$-lasso~$\frakM$ from a $1$-lasso~$\frakL$
by attaching $k$-lassos.
We will call this lasso~$\frakL$ the \emph{main lasso} of~$\frakM$.

The types we use for $k$-lassos are based on the same principles as those for simple
lassos,
but we have to nest them in order to take the branching of a hierarchical lasso
into account.
\begin{Def}
Let $t : \dom(t) \to C$ be a labelled tree and $m < \omega$.

(a) For a branch~$\beta$ of~$t$, we set
\begin{align*}
  \wtp_m(\beta) := [\sigma,\tau]\,,
\end{align*}
if $\beta$~has a factorisation of $m$-type $\langle\sigma,\tau\rangle$.
(By Proposition~\ref{Prop: types of factorisations}, this is well-defined.)

(b) For $k < \omega$, we define
\begin{align*}
  \tp^0_m(t) &:= \bigset{ \wtp_m(\beta) }{ \beta \text{ a branch of } t }\,, \\
  \tp^{k+1}_m(t) &:= \tp^0_m(\TP^k_m(t))\,,
\end{align*}
where $\TP^k_m(t) : T \to C \times \PSet(\Theta^k_m)$ is the tree with labelling
\begin{align*}
  &\TP^k_m(t)(v) := \bigl\langle t(v),\ 
                      \set{ \tp^k_m(t|_u) }{ u \text{ a successor of } v }\bigr\rangle\,.
\end{align*}
($t(v)$~is the label of the vertex~$v$ and $t|_u$ denotes the subtree attached to~$u$.)
\markenddef
\end{Def}

We will prove that the functions~$\tp^k_m$ form a family of type functions.
Note that it follows immediately from the definition that they satisfy
Properties \textsc{(S2)} and \textsc{(S3).}
\begin{Lem}
\itm{(a)}
Let $\frakM$~be a $k$-lasso and $\frakN$~a $k'$-lasso. Then
\begin{align*}
  \calU(\frakM) \sim \calU(\frakN)
  \qtextq{implies}
  \tp^k_m(\frakM) = \tp^k_m(\frakN)\,.
\end{align*}

\itm{(b)} For every type~$\tau$, there exists an $\MSO$-formula~$\varphi$ such that
\begin{align*}
  \calU(\frakM) \models \varphi \quad\iff\quad \tp^k_m(\frakM) = \tau\,.
\end{align*}
\end{Lem}

Thus, to prove that the class of $k$-lassos has the unravelling property
it is sufficient to show that $\tp^k_m$ also satisfies Property~\textsc{(S1).}
We will do so by induction on~$k$. The base case of this induction rests
on the following lemma.
\begin{Lem}\label{Lem: equivalent to 1-lasso}
Let $\calL_k$~be the class of all $k$-lassos and
let $\frakM$~be a $k$-lasso such that, for every vertex~$v$
and all branches $\beta$~and~$\gamma$ starting at a successor of~$v$,
we have $\wtp_m(\beta) = \wtp_m(\gamma)$.
Then $\frakM \simeq^m_{\calL_k} \frakN$, for some $1$-lasso~$\frakN$.
\end{Lem}
\begin{proof}
We prove the claim by induction on~$k$.
For $k = 1$, we can take $\frakN := \frakM$.
Hence, suppose that $k > 1$.
By inductive hypothesis, every sublasso attached to the main lasso is equivalent
to some $1$-lasso. Replacing them by these $1$-lassos, we may assume that $k = 2$.

We start by getting rid of the sublassos attached to the main loop of~$\frakM$.
Fix a vertex~$v$ on the main loop of~$\frakM$ and let $\frakP$~be the cycle from~$v$ back
to~$v$.
Let $\frakL$~be a sublasso attached to~$v$.
By Lemma~\ref{Lem: lasso type determines m-L-equivalence}, we have
$\frakL \simeq^m_{\calL_1} \frakP$.
Hence, we can replace~$\frakL$ by~$\frakP$.
Let $\frakM'$~be the $2$-lasso obtained by these substitutions, let $\frakK'$~be
the main loop of~$\frakM'$ (including all the sublassos), and let $\frakK''$~be
the loop obtained from~$\frakK'$ by removing the sublassos.
As every sublasso attached to the main loop~$\frakK'$ is isomorphic to~$\frakK''$,
it follows that
$\frakK' \sim \frakK''$.
Let $\frakM''$~be the $2$-lasso obtained from~$\frakM'$ by replacing the loop~$\frakK'$
by~$\frakK''$. Then
\begin{align*}
  \frakM'' \sim \frakM' \simeq^m_{\calL_1} \frakM\,.
\end{align*}

It remains to remove the sublassos of~$\frakM''$ attached to the tail.
We prove the claim by induction on the number of vertices of~$\frakM''$
that have sublassos attached.
If there are none, we are done. Otherwise, let $v$~be the last such vertex,
let $\frakL$~be the part of the main lasso that is attached to~$v$
and let $\frakK$~be some sublasso attached to~$v$.
By Lemma~\ref{Lem: lasso type determines m-L-equivalence}, we have
$\frakK \simeq^m_{\calL_1} \frakL$.
Let $\frakM'''$~be the $2$-lasso obtained from~$\frakM''$
by replacing all sublassos attached to~$v$ by a copy of~$\frakL$
and let $\frakM^{(4)}$ be the $2$-lasso obtained by removing all these sublassos.
Then
\begin{align*}
  \frakM^{(4)} \sim \frakM''' \simeq^m_{\calL_2} \frakM''\,.
\end{align*}
As $\frakM^{(4)}$~has one less vertex with sublassos attached,
we can use the inductive hypothesis to find an $1$-lasso~$\frakN$ with
$\frakN \simeq^m_{\calL_2} \frakM^{(4)} \simeq^m_{\calL_2} \frakM''
        \simeq^m_{\calL_2} \frakM$.
\end{proof}

\begin{Prop}\label{Prop: tpkm implies simeq}
Let $\frakM$~be a $k$-lasso and $\frakN$~a $k'$-lasso. For $m \geq 1$,
\begin{align*}
  \tp^k_m(\frakM) = \tp^k_m(\frakN)
  \qtextq{implies}
  \frakM \simeq^m_{\calL_K} \frakN\,,
\end{align*}
where $\calL_K$~is the class of all $K$-lassos with $K :=\max(k,k')$.
\end{Prop}
\begin{proof}
We prove the claim by induction on~$k$.
First, suppose that $k = 1$.
Then $\tp^1_m(\frakM) = \tp^1_m(\frakN)$ and $m \geq 1$ implies that $\frakN$~satisfies
the conditions of Lemma~\ref{Lem: equivalent to 1-lasso} (since $\frakM$~does).
Therefore, we can find some $1$-lasso~$\frakN'$ with $\frakN' \simeq^m_{\calL_K} \frakN$.
As~$\tp^1_m(\frakM)$ determines $\wtp_m(\beta)$,
where $\beta$~is the unique branch of~$\calU(\frakM)$,
it follows by Lemma~\ref{Lem: lasso type determines m-L-equivalence} that
$\frakM \simeq^m_{\calL_K} \frakN' \simeq^m_{\calL_K} \frakN$.

For the inductive step, suppose that $k > 1$.
Let $\beta$~and~$\gamma$ be the branches of $\TP^{k-1}_m(\calU(\frakM))$
and $\TP^{k-1}_m(\calU(\frakN))$ that correspond to their main lassos.

We first consider the case where $\wtp_m(\beta) = \wtp_m(\gamma)$.
For every $\tp^{k-1}_m$-type~$\sigma$, we pick a representative~$\frakC_\sigma$.
Let $\frakM'$~and~$\frakN'$ be the $k$-lassos obtained by replacing every sublasso of
type~$\sigma$ by its representative~$\frakC_\sigma$.
By inductive hypothesis and Proposition~\ref{Prop: replacing attached subsystems},
it follows that
$\frakM \simeq^m_{\calL_K} \frakM'$ and $\frakN \simeq^m_{\calL_K} \frakN'$.
As the $m$-types of $\beta$~and~$\gamma$ are conjugate
(including all the information about attached sublassos),
it follows by Lemma~\ref{Lem: lasso type determines m-L-equivalence} that the two
lassos $\frakA$~and~$\frakB$ that correspond to the branches $\beta$~and~$\gamma$ are
$\simeq^m_\calL$-equivalent, even with the additional labelling provided by
$\TP^{k-1}_m$.
Note that $\frakM'$~is the $k$-lasso obtained from~$\frakA$
by attaching all representatives $\frakC_\sigma$ as indicated by this labelling,
and $\frakN'$~is obtained from~$\frakB$ in the same way.
By Proposition~\ref{Prop: replacing attached subsystems} it therefore follows that
$\frakM' \simeq^m_{\calL_K} \frakN'$.
Consequently,
\begin{align*}
  \frakM \simeq^m_{\calL_K} \frakM' \simeq^m_{\calL_K} \frakN' \simeq^m_{\calL_K} \frakN\,.
\end{align*}

It remains to consider the case where $\beta$~and~$\gamma$ have different $m$-types.
As $\frakM$~and~$\frakN$ have the same type, there exists a branch~$\gamma'$
of $\TP^{k-1}_m(\calU(\frakN))$ whose $m$-type is conjugate to that of~$\beta$.
We will construct a $(k-1)$-lasso $\frakN' \simeq^m_{\calL_K} \frakN$ such that
$\tp^k_m(\frakN') = \tp^k_m(\frakM)$ and the main lasso of~$\frakN'$ has the same type
as~$\gamma'$. Then the claim follows from the special case proved above.

We construct~$\frakN'$ by choosing a copy of~$\gamma'$ as its main lasso.
For every successor~$u$ of a vertex~$v$ of~$\gamma'$ that does not itself belong
to~$\gamma'$, we attach a copy of~$\frakC_\sigma$ to the corresponding vertex of~$\frakN'$,
where $\sigma$~is the type of the sublasso of~$\frakN$ rooted at~$u$.
By the definition of~$\tp^k_m$ it follows that
\begin{align*}
  \tp^k_m(\frakN') = \tp^k_m(\frakN) = \tp^k_m(\frakM)\,,
\end{align*}
as desired. Furthermore, Proposition~\ref{Prop: replacing attached subsystems} implies
that $\frakN' \simeq^m_{\calL_K} \frakN$.
\end{proof}

Using Theorem~\ref{Thm: type functions imply unravelling} we now immediately obtain
the following statement.
\begin{Thm}\label{Thm: k-lassos have unravelling property}
For every~$k$, the class of all $k$-lassos has the unravelling property.
\end{Thm}

\section{Reductions}   %%%%%%%%%%%%%%%%%%%%%%%%%%%%%%%%%%%%%%%%%%%%%%%%%%%%%%%%%%%%%%%%%%%
\label{Sect: reductions}

We would like to define reductions that allow us to prove that a certain class
has the unravelling property when we already know that
some other class has this property.
To do so, we encode every transition system of the first class by some
system in the second one.
The main example we will be working with is a function~$\varrho$
that removes certain attached subsystems and uses additional vertex labels to remember
the $m$-theories of all deleted system.
Up to equivalence of $m$-theories, we can undo this operation by a
function~$\eta$ that attaches to each vertex labelled by some
$m$-theory~$\theta$ some fixed system with theory~$\theta$.
Let us give a general definition of such pairs of maps.

\begin{Def}
Let $\calC$~and~$\calD$ be classes of transition systems and $k,m < \omega$.
A~function $\varrho : \calC \to \calD$ is a \emph{$(k,m)$-encoding map}
if there exists a function $\eta : \calD \to \calC$ such that
\begin{enumerate}[label={\normalfont\scshape(E\arabic*)}]
\item $\varrho(\eta(\frakD)) \simeq^k_\calD \frakD\,,
  \quad\text{for all } \frakD \in \calD\,.$
\item $\varrho(\frakC) \simeq^k_\calD \varrho(\frakC') \qtextq{implies}
       \frakC \simeq^m_\calC \frakC'\,,
  \quad\text{for all } \frakC,\frakC' \in \calC\,.$
\end{enumerate}
In this case, we call the function~$\eta$ a \emph{$(k,m)$-decoding map}
for~$\varrho$.
\markenddef
\end{Def}

\begin{Exam}
Let $\calT$~be the class of all trees and $\calC \supseteq \calT$ any class containing it.
The unravelling operation $\calU : \calC \to \calT$ is an $(m,m)$-encoding map
and the identity function $\mathrm{id} : \calT \to \calC$ the corresponding $(m,m)$-decoding map.
For \textsc{(E1),} it is sufficient to note that
$\calU(\mathrm{id}(\frakT)) = \frakT$, for every tree~$\frakT$.
For \textsc{(E2),} consider two systems $\frakS,\frakS' \in \calC$.
Then
\begin{align*}
  \calU(\frakS) \simeq^m_\calT \calU(\frakS')
  \qtextq{implies}
  \frakS \sim \calU(\frakS) \simeq^m_\calC \calU(\frakS') \sim \frakS'\,.
\end{align*}
\end{Exam}

Let us note that the two axioms of an encoding map imply dual axioms
with the functions $\varrho$~and~$\eta$ exchanged.
\begin{Lem}\label{Lem: e3 and e4}
Let $\eta : \calD \to \calC$ be a $(k,m)$-decoding map for
$\varrho : \calC \to \calD$.
\begin{enumerate}[label={\normalfont\scshape(E\arabic*)}, start=3]
\item $\eta(\varrho(\frakC)) \simeq^m_\calC \frakC\,,
  \quad\text{for all\/ } \frakC \in \calC\,.$
\item $\frakD \simeq^k_\calD \frakD' \qtextq{implies}
  \eta(\frakD) \simeq^m_\calC \eta(\frakD')\,,
  \quad\text{for all\/ } \frakD,\frakD' \in \calD\,.$
\end{enumerate}
\end{Lem}
\begin{proof}
\textsc{(E3)} By \textsc{(E1)} and \textsc{(E2),}
\begin{align*}
  \varrho(\eta(\varrho(\frakC))) \simeq^k_\calD \varrho(\frakC)
  \qtextq{implies}
  \eta(\varrho(\frakC)) \simeq^m_\calC \frakC\,.
\end{align*}

\noindent
\textsc{(E4)} By \textsc{(E1)} and \textsc{(E2),}
\begin{align*}
  \varrho(\eta(\frakD))
    \simeq^k_\calD \frakD
    \simeq^k_\calD \frakD'
    \simeq^k_\calD \varrho(\eta(\frakD'))
  \qtextq{implies}
  \eta(\frakD) \simeq^m_\calC \eta(\frakD')\,.
\end{align*}
\upqed
\end{proof}

The axioms of an encoding map were chosen to guarantee the property
stated in the following lemma.
It will be used below to prove that encoding maps can be used to transfer
the unravelling property from one class to another.
\begin{Lem}\label{Lem: bisimulation-invariant decoding}
Let $\varrho : \calC \to \calD$ a $(k,m)$-encoding map
and $\eta : \calD \to \calC$ a $(k,m)$-decoding map for~$\varrho$.
For every $\MSO$-formula~$\varphi$ of quan\-ti\-fier-rank~$m$ that is
bisim\-u\-la\-tion-invari\-ant over~$\calC$, there exists an
$\MSO$-formula~$\hat\varphi$ of quan\-ti\-fier-rank $k$ that is
bisim\-u\-la\-tion-invari\-ant over~$\calD$ such that
\begin{align*}
  \frakC \models \varphi \quad\iff\quad \varrho(\frakC) \models \hat\varphi\,,
  \quad\text{for all } \frakC \in \calC\,.
\end{align*}
\end{Lem}
\begin{proof}
By \textsc{(E2)} and Proposition~\ref{Prop: simeq and bisim. theories},
\begin{alignat*}{-1}
  \varrho(\frakC) \equiv^k_\calD \varrho(\frakC')
  &\quad\Rightarrow\quad
  \varrho(\frakC) \simeq^k_\calD \varrho(\frakC') \\
  &\quad\Rightarrow\quad
  \frakC \simeq^m_\calC \frakC'
  \quad\Rightarrow\quad
  \frakC \equiv^m_\calC \frakC'\,.
\end{alignat*}
Hence, there exists a function~$h$ on $\MSO$-theories such that
\begin{align*}
  \Th^m_\calC(\frakC) = h\bigl(\Th^k_\calD(\varrho(\frakC))\bigr)\,.
\end{align*}
We set
\begin{align*}
  \hat\varphi := \Lor h^{-1}[\Theta_\varphi]\,,
\end{align*}
where $\Theta_\varphi$~is the set of all $\MSO_m$-theories containing~$\varphi$.
Note that $\hat\varphi$~is bisim\-u\-la\-tion-invari\-ant over~$\calD$
since bisim\-u\-la\-tion-invari\-ant formulae are closed under boolean operations.
Furthermore, $\hat\varphi$~has quan\-ti\-fier-rank~$k$ and
\begin{align*}
  \varrho(\frakC) \models \hat\varphi
  &\quad\iff\quad
    h\bigl(\Th_k(\varrho(\frakC))\bigr) \in \Theta_\varphi \\
  &\quad\iff\quad
    \varphi \in h\bigl(\Th_k(\varrho(\frakC))\bigr)
      = \Th^m_\calC(\frakC)
   \quad\iff\quad \frakC \models \varphi\,.
\end{align*}
\upqed
\end{proof}

It remains to show how to use encoding maps to transfer the unravelling property.
Just the existence of such a map is not sufficient. It also has to be what
we call definable.
\begin{Def}
Let $\calC$ be a class of transition systems.

(a) A $(k,m)$-encoding map $\varrho : \calC \to \calD$ is \emph{definable} if,
for every $\MSO$-formula~$\varphi$ that is bisim\-u\-la\-tion-invari\-ant
over trees, there exists an $\MSO$-formula~$\hat\varphi$ that is
bisim\-u\-la\-tion-invari\-ant over trees such that
\begin{align*}
  \calU(\varrho(\frakC)) \models \varphi
  \quad\iff\quad
  \calU(\frakC) \models \hat\varphi\,,
  \quad\text{for all } \frakC \in \calC\,.
\end{align*}

(b) We say that $\calC$~is \emph{reducible} to a family~$(\calD_m)_{m<\omega}$
of classes if there exist a map $g : \omega \to \omega$ and, for each $m < \omega$,
functions $\varrho_m : \calC \to \calD_m$ and $\eta_m : \calD_m \to \calC$
such that $\varrho_m$~is a definable $(g(m),m)$-encoding map and
$\eta_m$~a corresponding $(g(m),m)$-decoding map.
\markenddef
\end{Def}
\noindent
(The only reason why we use a family of classes to reduce to, instead of a single one
is so that we can have the labellings of systems in~$\calD_m$ depend on the
quantifier-rank~$m$.)

\begin{Thm}\label{Thm: reducing the unravelling property}
Suppose that $\calC$~is reducible to~$(\calD_m)_{m<\omega}$.
If every class~$\calD_m$ has the unravelling property, so does~$\calC$.
\end{Thm}
\begin{proof}
Let $\varphi$~be bisim\-u\-la\-tion-invari\-ant over~$\calC$ and let $m$~be its
quan\-ti\-fier-rank.
By Lemma~\ref{Lem: bisimulation-invariant decoding}, there exists an
$\MSO$-formula~$\psi$ that is bisim\-u\-la\-tion-invari\-ant over~$\calD_m$
such that
\begin{align*}
  \frakC \models \varphi \quad\iff\quad \varrho_m(\frakC) \models \psi\,,
  \quad\text{for all } \frakC \in \calC\,.
\end{align*}
Using the unravelling property of~$\calD_m$,
we can find an $\MSO$-formula~$\hat\psi$ that is bisim\-u\-la\-tion-invari\-ant
over trees such that
\begin{align*}
  \frakD \models \psi \quad\iff\quad \calU(\frakD) \models \hat\psi\,,
  \quad\text{for all } \frakD \in \calD_m\,.
\end{align*}
Finally, definability of~$\varrho_m$ provides an $\MSO$-formula~$\hat\varphi$
that is bisim\-u\-la\-tion-invari\-ant over trees such that
\begin{align*}
  \calU(\varrho_m(\frakC)) \models \hat\psi
  \quad\iff\quad
  \calU(\frakC) \models \hat\varphi\,,
  \quad\text{for all } \frakC \in \calC\,.
\end{align*}
Consequently, we have $\frakC \models \varphi$ if, and only if,
$\calU(\frakC) \models \hat\varphi$, for all $\frakC \in \calC$.
\end{proof}

\section{Finite Cantor--Bendixson rank}   %%%%%%%%%%%%%%%%%%%%%%%%%%%%%%%%%%%%%%%%%%%%%%%%
\label{Sect: CB rank}

One common property of $k$-lassos is that the trees we obtain by unravelling them
all have finite Cantor--Bendixson rank.
In this section we will generalise our results to cover transition systems with
this more general property.
The proof below consists in a two-step reduction to the class of $k$-lassos.
\begin{Def}
Let $\frakT$~be a finitely branching tree.
The \emph{Cantor--Bendixson derivative} of~$\frakT$
is the tree~$\frakT'$ obtained from~$\frakT$ by removing all subtrees that have only
finitely many infinite branches.
The \emph{Cantor-Bendixson rank} of a tree~$\frakT$ is the least ordinal~$\alpha$ such
that applying $\alpha+1$ Cantor--Bendixson derivatives to~$\frakT$ results in an empty
tree.
The \emph{Cantor--Bendixson rank} of a transition system~$\frakS$ is equal to the
Cantor--Bendixson rank of its unravelling.
\markenddef
\end{Def}

We can go from the class of $k$-lassos to that of
systems with bounded Cantor--Bendixson rank in two steps.
\begin{Def}
(a) A transition system is a \emph{generalised $k$-lasso} if it is obtained from a finite
tree by attaching (one or several) $k$-lassos to every leaf.

(b) A transition system~$\frakT$ is a \emph{tree extension} of~$\frakS$
if $\frakT$~is obtained from~$\frakS$ by attaching an arbitrary number of finite
trees to some of the vertices.
\markenddef
\end{Def}

With these two notions we can characterise the property of having
bounded Cantor--Bendixson rank as follows.
\begin{Prop}\label{Prop: CB-rank and k-lassos}
Let $\frakS$~be a finite transition system.

\noindent
\itm{(a)} For every $k < \omega$, the following statements are equivalent.
\begin{enum1}
\item $\frakS$~has Cantor--Bendixson rank at most~$k$.
\item $\frakS$~is bisimilar to a tree extension of a generalised $(k+1)$-lasso.
\end{enum1}

\noindent
\itm{(b)} The following statements are equivalent.
\begin{enum1}
\item $\frakS$~has finite Cantor--Bendixson rank.
\item $\frakS$~is bisimilar to a tree extension of a generalised $k$-lasso,
  for some $k < \omega$.
\item Every strongly connected component of~$\frakS$ is either a singleton or a cycle.
\end{enum1}
\end{Prop}
\begin{proof}
(a) follows by induction on~$k$.
For $k=0$, note that a transition system~$\frakS$ has Cantor--Bendixson rank~$0$
if, and only if, its unravelling consists of finitely many infinite branches
and attached finite subtrees.
This is the case if, and only if, $\frakS$~is bisimilar to a tree extension of
a generalised $1$-lasso.

For $k > 0$, note that $\frakS$~has Cantor--Bendixson rank at most~$k$
if, and only if, in its unravelling we can choose finitely many branches such that
all subtrees that do not contain any of them have Cantor--Bendixson rank at most~$k-1$.
By inductive hypothesis, this is the case if, and only if, the unravelling
is bisimilar to a tree with finitely many infinite branches to which tree extensions
of generalised $k$-lassos are attached at arbitrary vertices.
Such a structure is bisimilar to a tree extension of a generalised $(k+1)$-lasso.

(b)
(1)~$\Leftrightarrow$~(2)
follows by~(a).

(3)~$\Rightarrow$~(2)
Suppose that every strongly connected component of~$\frakS$ is either a singleton or a
cycle.
In the partial order formed by all strongly connected components of~$\frakS$
(ordered by the reachability relation), fix a chain of maximal length that
consists only of components that are cycles and let $k$~be its length.
By induction on~$k$ it follows that
we can partially unravel~$\frakS$ into a tree extension of a generalised $k$-lasso.

(1)~$\Rightarrow$~(3)
Suppose that $\frakS$~has a strongly connected component that is not a cycle nor a
singleton. This component contains a state~$s$ with two distinct paths $u$~and~$v$
from~$s$ back to~$s$. (These paths may share vertices.)
Consequently, the unravelling of~$\frakS$ contains a copy $\{u,v\}^*$ of the complete
binary tree. In particular, it has infinite Cantor--Bendixson rank.
\end{proof}

To prove the unravelling property for the transition systems of bounded Cantor--Bendix\-son
rank, we proceed in two steps. First we consider generalised $k$-lassos
and then their tree extensions.
\begin{Thm}\label{Thm: characterisation for generalised k-lassos}
For fixed~$k$, the class of all generalised $k$-lassos has the unravelling property.
\end{Thm}
\begin{proof}
We show that the class is reducible to a certain class of finite trees.
Let $\Theta^k_m$~be the set of all $\tp^k_m$-types.
It follows by Proposition~\ref{Prop: tpkm implies simeq} that the $\tp^k_m$-type
of a $k'$-lasso determines whether or not it is in fact a $k$-lasso.
Let $\Lambda^k_m \subseteq \Theta^k_m$ be the subset of all types that
correspond to $k$-lassos and
let $\calT^k_m$~be a certain class of finite trees labelled by subsets
of~$\Lambda^k_m$ that we will define below.

We start by defining an $(m,m)$-encoding map $\varrho_m : \calH_k \to \calT^k_m$
as follows.
Given a generalised $k$-lasso~$\frakM$, $\varrho_m(\frakM)$~is the finite tree obtained
from the unravelling~$\calU(\frakM)$ by removing all subtrees whose type belongs
to~$\Lambda^k_m$. We label each vertex~$v$ by the set of all types belonging
to one of the removed subtrees attached to~$v$.
To define the corresponding $(m,m)$-decoding map $\eta_m : \calT^k_m \to \calH_k$
we fix, for every $\tau \in \Lambda^k_m$ some $k$-lasso~$\frakC_\tau$ of type~$\tau$.
Given a labelled tree~$\frakT$ the map~$\eta_m$
attaches to every vertex with label $\{\tau_0,\dots,\tau_{n-1}\}$
copies of~$\frakC_{\tau_0},\dots,\frakC_{\tau_{n-1}}$.
Finally, we chose for $\calT^k_m$~the image of the map~$\varrho_m$.

We claim that the maps $\varrho_m$~and~$\eta_m$ form a definable family of
encoding and decoding maps. There are three conditions to check.

\textsc{(E1)} By definition, $\varrho_m(\eta_m(\frakT)) = \frakT$,
for every tree~$\frakT$.
(We have to be careful to check that $\varrho_m$~does not remove more vertices than those
added by~$\eta_m$. But this cannot happen, as $\frakT \in \calT^k_m$, i.e.,
$\frakT$~is of the form $\varrho_m(\frakM)$, for some~$\frakM$.)

\textsc{(E2)} Let $\frakM$~and~$\frakN$ be generalised $k$-lassos with
$\varrho_m(\frakM) \simeq^m_{\calT^k_m} \varrho_m(\frakN)$.
Then there exists a finite sequence $\frakT_0,\dots,\frakT_n$ of trees such that
\begin{align*}
  \frakT_0 = \varrho_m(\frakM)\,,\quad
  \frakT_n = \varrho_m(\frakN)\,,\quad\text{and}\quad
  \frakT_i \sim \frakT_{i+1} \text{ or } \frakT_i \equiv_m \frakT_{i+1}\,,
\end{align*}
for all $i < n$.
Set $\frakL_0 := \frakM$, $\frakL_n := \frakN$, and
$\frakL_i := \eta_m(\frakT_i)$, for $0 < i < n$.
Then it follows that
$\frakL_i \sim \frakL_{i+1}$ or $\frakL_i \equiv_m \frakL_{i+1}$, for all $i < n$.
Consequently, $\frakM \simeq^m_{\calH_k} \frakN$.

(definability)
Note that $\varrho_m(\frakM)$ is a subtree of $\calU(\frakM)$.
Since the $\tp^k_m$-type of a subtree is definable in monadic second-order logic,
there exists an $\MSO$-formula~$\psi(x)$ defining $\varrho_m(\frakM)$
inside of~$\calU(\frakM)$.
Given an $\MSO$-formula~$\varphi$ we can therefore use the formula~$\psi$ to construct
a new $\MSO$-formula~$\hat\varphi$ such that
\begin{align*}
  \varrho_m(\frakM) \models \varphi \quad\iff\quad \calU(\frakM) \models \hat\varphi\,.
\end{align*}
Furthermore, if $\varphi$~is bisim\-u\-la\-tion-invari\-ant over the class of all trees,
so is~$\hat\varphi$.
\end{proof}

Using this intermediate step, we obtain the following proof for transition systems
with bounded Cantor--Bendixson rank.
\begin{Thm}\label{Thm: characterisation for CB-rank}
The class of all finite transition systems of Cantor--Bendixson rank at most~$k$
has the unravelling property.
\end{Thm}
\begin{proof}
First note that according to Lemma~\ref{Lem: extending the unravelling property}
it is sufficient to prove that the class~$\calE_k$ of all tree extensions of
generalised $k$-lassos has the unravelling property.
Let $\calH^m_k$~be the class of all generalised $k$-lassos where the vertices
are labelled by sets of $m$-theories.

To do so, we present a reduction to the class of generalised $k$-lassos.
Our $(m,m)$-encoding maps $\varrho_m : \calE_k \to \calH^m_k$ map
a tree extension~$\frakM$ to the generalised $k$-lasso $\varrho_m(\frakM)$
obtained by removing all attached finite trees.
To remember what was deleted, we label every vertex~$v$ with the set of
$m$-theories of the subtrees that were attached to~$v$.
The corresponding $(m,m)$-decoding map $\eta_m : \calH^m_k \to \calE_k$ simply adds a
representative of every $m$-theory to all vertices labelled by this theory.

To see that $\varrho_m$~and~$\eta_m$ form a definable family of encoding and decoding
maps, we have to check three conditions.

\textsc{(E1)}
We have $\varrho_m(\eta_m(\frakM)) = \frakM$, for every generalised $k$-lasso~$\frakM$.

\textsc{(E2)}
Suppose that $\varrho_m(\frakM) \simeq^m_{\calH^m_k} \varrho_m(\frakN)$.
As in the previous proof we can take a sequence of generalised $k$-lassos
witnessing this fact and modify it by reattaching the removed subtrees
to obtain a sequence witnessing that $\frakM \simeq^m_{\calE_k} \frakN$.

(definability)
As the $m$-theory of a subtree is definable in $\MSO$, we can construct an $\MSO$-formula
$\psi(x)$ defining $\varrho_m(\frakM)$ inside of~$\frakM$.
This formula can be used to define $\calU(\varrho_m(\frakM))$ inside~$\calU(\frakM)$.
\end{proof}

\begin{Cor}
Over the class of all finite transition systems with Cantor--Bendixson rank at most~$k$,
bisim\-u\-la\-tion-invari\-ant $\MSO$ coincides with~$\Lmu$.
\end{Cor}

\section{Conclusion}   %%%%%%%%%%%%%%%%%%%%%%%%%%%%%%%%%%%%%%%%%%%%%%%%%%%%%%%%%%%%%%%%%%%

We have shown in several simple examples how to characterise
bisim\-u\-la\-tion-invari\-ant $\MSO$ in the finite.
In particular, we have proved that it coincides with~$\Lmu$ over
\begin{itemize*}
\item every finite class (Theorem~\ref{Thm: finite classes}),
\item the class of all finite trees (Theorem~\ref{Thm: finite trees}),
\item the classes of all lassos, $k$-lassos, and generalised $k$-lassos
  (Theorems \ref{Thm: characterisation for lassos},
   \ref{Thm: k-lassos have unravelling property}, and
   \ref{Thm: characterisation for generalised k-lassos}),
\item the class of all systems of Cantor--Bendixson rank at most~$k$
  (Theorem~\ref{Thm: characterisation for CB-rank}).
\end{itemize*}
Our main tool in these proofs was the unravelling property
(Theorem~\ref{Thm: unvarelling property}).
It will be interesting to see how far our methods can be extended to more
complicated classes.
For instance, can they be used to prove the following conjecture\??
\begin{VarLem}[Conjecture]
If a class~$\calC$ of transition systems has the unravelling property,
then so does the class of all subdivisions of systems in~$\calC$.
\end{VarLem}
\noindent
A good first step seems to be the class of all finite transition systems
that have Cantor--Bendixson rank~$k$, for some $k < \omega$ that is not fixed.

In this paper we have considered only transition systems made out of paths
with very limited branching. To extend our techniques to classes allowing
for more branching seems to require new ideas.
A simple test case that looks promising is the class of systems with a
`lasso-decomposition' of width~$k$, i.e.,
something like a tree decomposition but where the pieces are indexed
by a lasso instead of a tree.

{\small

{\small
\begin{thebibliography}{10}

\bibitem{AndrekaVBeNe98}
{\sc H.~Andr\'eka, J.~v. Benthem, and I.~N\'emeti}, {\em {Modal languages and
  bounded fragments of predicate logic}}, Journal of Philosophical Logic, 27
  (1998), pp.~217--274.

\bibitem{vanBenthem76}
{\sc J.~v. Benthem}, {\em {Modal Correspondence Theory}}, {Ph.\,D.\ Thesis},
  {University of Amsterdam}, Amsterdam, 1976.

\bibitem{BlumensathColcombetLoeding07}
{\sc A.~Blumensath, T.~Colcombet, and C.~{L\"oding}}, {\em {Logical Theories
  and Compatible Operations}}, in {Logic and Automata: History and
  Perspectives}, J.~Flum, E.~{Gr\"adel}, and T.~Wilke, eds., Amsterdam
  University Press, 2007, pp.~73--106.

\bibitem{Carreiro15}
{\sc F.~Carreiro}, {\em {Fragments of Fixpoint Logics}}, {PhD Thesis},
  {Institute for Logic, Language and Computation}, Amsterdam, 2015.

\bibitem{CiardelliOtto17}
{\sc I.~Ciardelli and M.~Otto}, {\em {Bisimulation in Inquisitive Modal
  Logic}}, in Proc.\ 16th Conference on Theoretical Aspects of Rationality and
  Knowledge, \smaller{TARK} 2017, 2017, pp.~151--166.

\bibitem{DawarJanin04}
{\sc A.~Dawar and D.~Janin}, {\em {On the bisimulation invariant fragment of
  monadic $\Sigma_1$ in the finite}}, in {Proc. of the 24th Int. Conf. on
  Foundations of Software Technology and Theoretical Computer Science,
  \smaller{FSTTCS} 2004}, 2004, pp.~224--236.

\bibitem{GraedelHirschOtto02}
{\sc E.~Gr\"adel, C.~Hirsch, and M.~Otto}, {\em {Back and Forth Between Guarded
  and Modal Logics}}, {ACM Transactions on Computational Logics},  (2002),
  pp.~418--463.

\bibitem{GraedelThomasWilke02}
{\sc E.~Gr\"adel, W.~Thomas, and T.~Wilke}, {\em {Automata, Logic, and Infinite
  Games}}, {\smaller{LNCS} 2500}, Springer-Verlag, 2002.

\bibitem{Hirsch02}
{\sc C.~Hirsch}, {\em {Guarded Logics: Algorithms and Bisimulation}}, {Ph.\,D.\
  Thesis}, {\smaller{RWTH} Aachen}, Aachen, 2002.

\bibitem{JaninWalukiewicz96}
{\sc D.~Janin and I.~Walukiewicz}, {\em On the expressive completeness of the
  propositional mu-calculus with respect to monadic second order logic}, in
  Proc. of the 7th International Conference on Concurrency Theory,
  \smaller{CONCUR} 1996, 1996, pp.~263--277.

\bibitem{Makowsky04}
{\sc J.~A. Makowsky}, {\em {Algorithmic aspects of the Feferman-Vaught
  Theorem}}, Annals of Pure and Applied Logic, 126 (2004), pp.~159--213.

\bibitem{MollerRabinovitch99}
{\sc F.~Moller and A.~Rabinovitch}, {\em {On the expressive power of CTL*}}, in
  {Proc.\ 14th \smaller{IEEE} Symp.\ on Logic in Computer Science,
  \smaller{LICS}}, 1999, pp.~360--369.

\bibitem{MollerRabinovitch03}
\leavevmode\vrule height 2pt depth -1.6pt width 23pt, {\em {Counting on CTL*:
  on the expressive power of monadic path logic}}, Information and Computation,
  184 (2003), pp.~147--159.

\bibitem{PerrinPin04}
{\sc D.~Perrin and J.-{\'E}. Pin}, {\em {Infinite Words -- Automata,
  Semigroups, Logic and Games}}, Elsevier, 2004.

\bibitem{Rosen97}
{\sc E.~Rosen}, {\em {Modal logic over finite structures}}, Journal of Logic,
  Language and Information, 6 (1997), pp.~427--439.

\bibitem{Shelah75}
{\sc S.~Shelah}, {\em {The Monadic Second Order Theory of Order}}, Annals of
  Mathematics, 102 (1975), pp.~379--419.

\bibitem{Stirling11}
{\sc C.~Stirling}, {\em {Bisimulation and logic}}, in {Advanced topics in
  Bisimulation and Coinduction}, D.~Sangiorgi and J.~Rutten, eds., Cambridge
  University Press, 2011, pp.~172--196.

\end{thebibliography}}
\end{document}